\documentclass[aps,prl,twocolumn,superscriptaddress,showpacs]{revtex4}

\usepackage{graphicx}
\usepackage{graphicx}% Include figure files
\usepackage{dcolumn}% Align table columns on decimal point
\usepackage{color}
\usepackage{mathrsfs}%\mathscr
\usepackage[breaklinks,colorlinks = true,linkcolor = blue,urlcolor=blue,citecolor=blue]{hyperref}
\usepackage{amsmath}
\usepackage{soul}

\begin{document}

\title{Intrinsic and extrinsic non-stationary field-driven processes in spin 
ice}

\author{S. Erfanifam}
\affiliation{Hochfeld-Magnetlabor Dresden, Helmholtz-Zentrum
  Dresden-Rossendorf, D-01314 Dresden, Germany} 

\author{S. Zherlitsyn}
\affiliation{Hochfeld-Magnetlabor Dresden, Helmholtz-Zentrum
  Dresden-Rossendorf, D-01314 Dresden, Germany} 

\author{J. Wosnitza}
\affiliation{Hochfeld-Magnetlabor Dresden, Helmholtz-Zentrum
  Dresden-Rossendorf, D-01314 Dresden, Germany} 

\author{R. Moessner}
\affiliation{Max-Planck Institut f\"ur Physik komplexer Systeme, D-01187
  Dresden, Germany} 

\author{O.A. Petrenko}
\affiliation{University of Warwick, Department of Physics, Coventry CV4 7AL UK}

\author{G. Balakrishnan}
\affiliation{University of Warwick, Department of Physics, Coventry CV4 7AL UK}

\author{A.A. Zvyagin}
\affiliation{Max-Planck Institut f\"ur Physik komplexer Systeme, D-01187
  Dresden, Germany} 
\affiliation{B.I.~Verkin Institute for Low Temperature Physics and Engineering
  of the National Academy of Sciences of Ukraine, Kharkov, 61103, Ukraine}

%%%%%%%%%%%%%%%%%%%%%%%%%%%%%%%%%%%%%%%%%%%%%%%%%%%%%%%%%%%%%%%%%%%%%%%%%%%%%%%

\begin{abstract}
Non-equilibrium processes are probed by ultrasound waves in the spin-ice
material Dy$_2$Ti$_2$O$_7$ at low temperatures. The sound velocity and the
sound attenuation exhibit a number of anomalies versus applied magnetic field
for temperatures below the ``freezing'' temperature of approximately
500~mK. These robust anomalies can be seen for longitudinal and transverse
acoustic modes for different field directions. The anomalies show a broad
hysteresis. Most notable are peaks in the sound velocity, which exhibit
two distinct regimes: an intrinsic (extrinsic) one in which the data collapse
for different sweep rates when plotted as function of field strength (time). We
discuss our observations in context of the emergent quasiparticles which govern
the low-temperature dynamics of the spin ice.  
\end{abstract}

\pacs{75.50.-y, 62.65.+k, 43.35.+d}

\date{\today}

\maketitle

%%%%%%%%%%%%%%%%%%%%%%%%%%%%%%%%%%%%%%%%%%%%%%%%%%%%%%%%%%

Systems with competing interactions are of a great interest in many-body
physics. For example, in magnetic systems, frustration manifests in a number
of interesting properties, distinct from the ones of ordinary magnets or spin
glasses \cite{2006_moess}. Due to their (thermodynamically) large ground-state
degeneracy, frustrated magnets often remain in an apparently disordered state
down to the lowest temperatures.

Among the best examples of frustrated magnetic systems are those with a
pyrochlore magnetic lattice, which consists of corner-shared tetrahedra with a
magnetic ion at each tetrahedron vertex. Dy$_2$Ti$_2$O$_7$ and
Ho$_2$Ti$_2$O$_7$ belong to the family of rare-earth titanates with the
pyrochlore lattice, which have attracted much interest in recent years because
of their unusual spin-ice ground state (the arrangement of magnetic moments
there is analogous to the proton configuration in water ice), deconfined
fractionalized excitations with magnetic Coulomb interactions (``magnetic
monopoles''), and dynamics
\cite{1997_harris,2001a_bramwell,2008_castelnovo,
2009_bramwell,2002_matsuhira,2009_kadowaki,2001_ryzhkin,2009_jaubert,2001_bramwell,
2005_fennell,2006_tabata,2007_matsuhira,2008_jaubert,2009_fennell,2009_morris,
2010_slobinsky,Schiffer_susc}. 
A broad range of experimental techniques have been brought to bear (e.g.,
magnetic, thermodynamic, and neutron-scattering measurements)
\cite{1997_harris,2001a_bramwell,2008_castelnovo,
2009_bramwell,2002_matsuhira,2009_kadowaki,2001_ryzhkin,2009_jaubert,2001_bramwell,
2005_fennell,2006_tabata,2007_matsuhira,2008_jaubert,2009_fennell,2009_morris,
2010_slobinsky,Schiffer_susc,2011_nakanishi,2011_petrenko}. 
The application of a uniform magnetic field is particularly rich (as it can act
as an effectively staggered field on account of the non-collinear easy axes,
\cite{1998_moessner}) and has produced a symmetry-sustaining Kasteleyn
transition \cite{2008_jaubert}, dimensional reduction to kagome ice
\cite{2002_matsuhira,2009_kadowaki,2006_tabata,2007_matsuhira} and a Zeeman equivalent of
deflagration \cite{2010_slobinsky}.

The aim of our work is to study the low-temperature, ($T < 1.5$~K) behavior of
Dy$_2$Ti$_2$O$_7$ to understand the role of spin-ice physics and monopoles in
the low-energy field-induced non-stationary processes in spin ices. We employ
ultrasound measurements as these are well known for their outstanding
sensitivity to probe various phase transitions and crossover phenomena
\cite{2005_luthi}. In our experiments, the phase-sensitive technique
\cite{2001_Wolf} was used to detect the relative change of the sound velocity,
$\Delta v(T,H)/v$, and the sound attenuation, $\Delta\alpha$, as a function of
temperature and external magnetic field.

The non-stationary processes observed in field sweeps in spin ice, have been
argued \cite{2010_slobinsky} to arise from an inability of the phonons to carry
away the Zeeman energy released by the flipping spins when thermal runaway is
triggered due to ``supercooling'' arising when the sparseness of defects at low
temperatures induces an exponentially slowdown of the dynamics
\cite{2001_ryzhkin,2009_jaubert,2010_slobinsky}.

Ultrasound, as a direct probe of the phononic degrees of freedom, thus provides
a natural and direct handle on this phenomenon. We find that the sound velocity
exhibits sets of well-defined spikes as the field is swept. We analyze the
shape of these peaks, which are highly asymmetric on account of the
fundamentally distinct non-equilibrium mechanisms involved: an  ``intrinsic''
rise followed by an ``extrinsic'' fall, evidenced by data collapse for various
combinations of peak positions and sweep rates: the former involves the
release of (Zeeman) energy from spin, the latter transfer of energy
out of the sample. In addition, for the case of an applied field in the [111]
direction, we find a very clear signature of the transition to saturated spin
ice, including the onset of hysteresis at the monopole ``liquid-gas'' transition \cite{2003_sakakibara}.

{\em Methods:} Single crystals of Dy$_2$Ti$_2$O$_7$ were grown under oxygen-gas flow by the floating-zone method in an infrared furnace \cite{1998_balak}. Dy$_2$Ti$_2$O$_7$ has
a cubic crystal structure with the space group $Fd\overline{3}m$. The
crystallographic orientations were controlled by the X-ray-diffraction Laue technique. The
lengths of the samples used in the experiments are 2.57~mm in [111], 3.41~mm in
[001], and 1.09~mm in [112] direction, respectively. Resonance LiNbO$_3$ or
wide-band PVDF-film (polyvinylidene fluoride film) piezoelectric transducers
glued on the polished surfaces of the sample are implemented to generate and
detect ultrasound waves in the frequency range of 60 - 110~MHz. Multiple
ultrasonic echoes were observed in the experiment due to ultrasound-wave
reflections from the two parallel polished sample surfaces. The sample was fixed in a sample holder
made out of brass in a ${^3}$He cryostat. A RuO$_2$ thermometer was attached directly to the sample. The
sample was in vacuum and the sample holder was thermally connected to the ${^3}$He
chamber. We have studied the $c_{11}$ ({\bf k}$\|${\bf u}$\|$[001]), $c_{L} =
(c_{11}+2c_{12}+4c_{44})/3$ ({\bf k}$\|${\bf u}$\|$[111]), and $c_{T} =
(c_{11}+c_{44}-c_{12})/3$ ({\bf k}$\|$[111], {\bf u}$\bot${\bf k}) acoustic
modes. Here {\bf k}, {\bf u} are the wave vector and polarization of the
acoustic wave, respectively, and $c_{ij}$ are the elastic moduli. The sound
velocity $v({\bf k},{\bf u})$ is related to the elastic modulus via $c=\rho
v^2$, where $\rho$ is the mass density of the crystal. We have also performed
ultrasound measurements for the [112] direction ({\bf k}$\|${\bf u}$\|$[112]). The applied
magnetic field was directed along the sound-propagation direction. Most of the
field dependences of the sound velocity and attenuation are reported for the
zero-field-cooled (ZFC) conditions.

\begin{figure}
\begin{center}
\vspace{-0.0cm}
\includegraphics[scale=0.33]{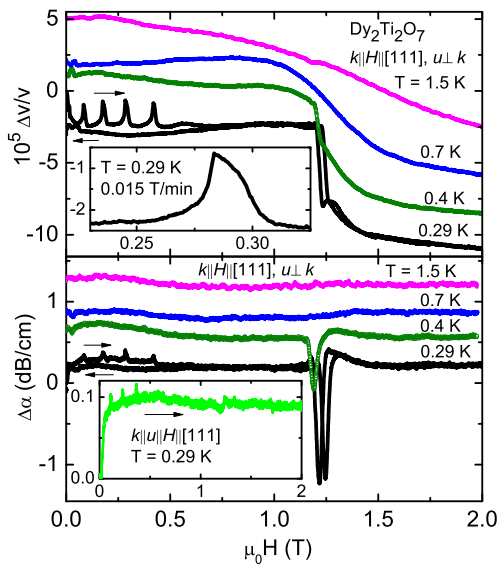}
\end{center}
\vspace{-0.5cm}
\caption{(Color online) Field dependence of the sound velocity, $\Delta v/v$,
  (top) and the sound attenuation, $\Delta\alpha$ (bottom) of the transverse
  ultrasonic wave propagating along the [111] ($k\|H\|$[111] and $u\bot k$)
  direction (acoustic $c_{T}$ mode) at different temperatures measured under ZFC condition. Results for up and down field-sweeps (indicated by arrows)
  are shown for the lowest temperature of 0.29~K. All data curves are arbitrarily shifted along the y-axis for clarity. Upper inset shows some
  details for a single peak in the sound velocity. The field dependence of the sound
  attenuation for the longitudinal mode (acoustic $c_{L}$ mode) at 0.29~K is
  given for comparison in the lower inset. Only the $c_{T}$ mode demonstrates a
  sharp attenuation anomaly at about 1.25~T.} 
\label{fig1}
\end{figure}

{\em Results:} Figure~1 shows typical field sweeps of the sound velocity and the
sound attenuation for the $c_{T}$ acoustic mode measured at various
temperatures. Several peaks appear in both the acoustic characteristics below
0.5~T, in the temperature range of 0.29 - 0.45~K ($c_{11}$ and $c_L$ acoustic
modes show similar features). In addition, there is an abrupt drop for the sound
velocity and an anomaly for the attenuation occurs at 1.25~T. It turns out that the
attenuation of the longitudinal mode $c_L$ (see the inset in the low panel of
Fig.~1) does not reveal any feature at 1.25~T, while the low-field peaks are
present. All features (except of the one at 1.25~T) disappear completely at
higher temperatures, $T \ge 0.5$~K.

In more detail, we have measured field dependences of the sound velocity using
sweep rates in the range of 0.015 - 0.15~T/min at $T$= 0.29~K (see
Fig.~2). This reveals a changeover from narrow peaks to broad peaks by
increasing the sweep rate. The maximal height of the peaks for various sweep
rates was quite comparable for all peaks. The number of peaks varies little 
for different field orientations and for different
sweep rates (Fig.~2). With increasing sweep-rate (and thus peak width), the
separation between the peaks decreases, so that peaks merge for large sweep
rates; also the positions of the maxima are shifted in field; a clear sign
for their non-equilibrium nature. A further indication that we are
not probing equilibrium spin configurations is provided by the thermometer
installed on the sample which shows a change of temperature at the values of
the field where the peaks of the sound velocity appear \cite{2010_slobinsky}
(inset of Fig.~2). Due to the nature of the thermometer's contact, the
absolute scale of the spikes is likely understated. It looks like the change of
sound velocity thus in large part arises from the change in the temperature of
the sample at these values of the magnetic field. The phonons thus provide an
in-situ thermometer! However, additional experimental and theoretical work is required in order to convert the sound-velocity change quantitatively into a temperature change.

\begin{figure}
\begin{center}
\vspace{-0.0cm}
\includegraphics[scale=0.33]{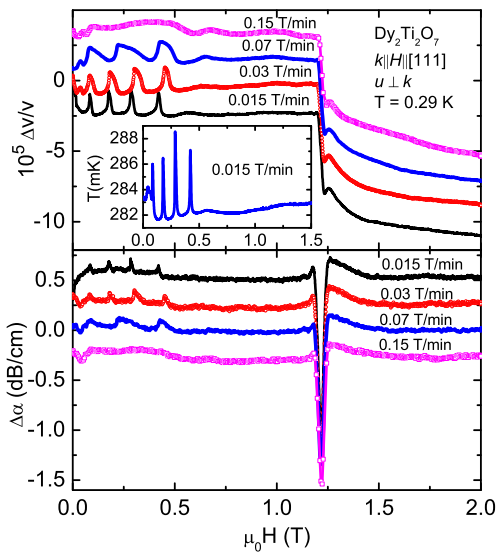}
\end{center}
\vspace{-0.5cm}
\caption{(Color online) The sound velocity (top) and the sound attenuation
  (bottom) of the $c_{T}$ acoustic mode measured at 0.29~K and various
  field-sweep rates. All data were obtained under ZFC condition. The data curves are arbitrarily shifted along the y-axis for clarity. The
  inset shows the temperature change, measured during an up field sweep by a RuO$_2$ thermometer directly
  attached to the sample.} 
\label{fig2}
\end{figure}

Before we analyze these findings in more detail, we briefly review the basic
model for spin ice \cite{2001a_bramwell}. To a good approximation, we are
dealing with Ising spins $\sigma = \pm 1$ whose magnetic moment of $|\pmb{\mu}|
\approx 10 \mu_B$ points along the local easy axis which joins the
ion site with the centres of the tetrahedra which share it. All states obeying
the ice rules (with two spins pointing into, and two out of, each tetrahedron)
are (near-)degenerate ground states.

Tetrahedra violating the ice rules appear as pointlike defects experiencing a
mutual relative {\em magnetic} Coulomb interaction $\frac{\mu_0 Q^2}{4 \pi r}$,
where $Q \approx 4.6 \mu_B/$\AA ~is the effective magnetic charge of these
monopoles. Single spin flips are only possible at low temperature when they
correspond to motion of monopoles -- otherwise they are suppressed by an
exponentially small Boltzmann factor ${\rm exp} [- \Delta/T]$, with $\Delta
\approx 4.3$~K. This leads to a strong slowdown of the dynamics below $T_f
\approx 500 - 600$~mK
\cite{2001_ryzhkin,2009_jaubert,2009_fennell,Schiffer_susc}.

As a result, at low $T$, field sweeps drive the system out of equilibrium: The
magnetization change demanded by the changing field cannot be established
sufficiently fast. Instead, ``reequilibration'' can occur in the form of
avalanches with the temperature rising up to $T_f$ \cite{2010_slobinsky}, which show up as the peaks in our
measurements of the sound velocity.

\begin{figure}
\begin{center}
\vspace{-0.0cm}
\includegraphics[scale=0.33]{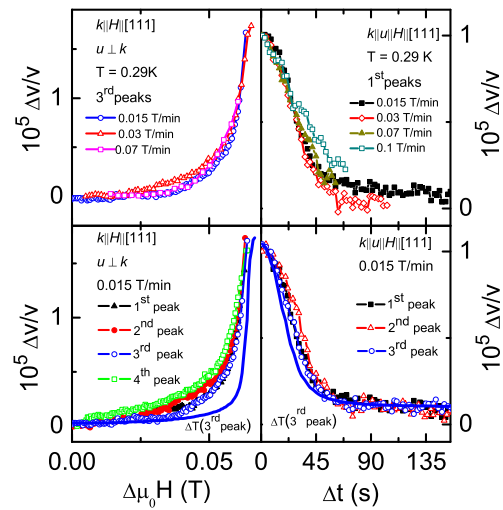}
\end{center}
\vspace{-0.5cm}
\caption{(Color online) Behavior of the rising parts (left) of the
  sound-velocity peaks of the $c_{T}$ acoustic mode for different sweep rates
  (upper panel) and different peaks (lower panel) as a function of magnetic field at $T$ = 0.29~K. The right panels show the descending
  parts of the velocity peaks for the $c_{L}$ acoustic mode as a
  function of time. The blue solid (lower) lines are uncalibrated temperature traces.} 
\label{compare1}
\end{figure}

With this in hand, let us analyze these peaks in more detail. As a starting
point, note that when the peaks are well developed (e.g., for the slow sweep
rate, 0.015~T/min), their overall shape is almost independent of the field
at which the peak is triggered (Fig. \ref{compare1}, bottom). Although we have observed the sound-velocity and the sound-attenuation peaks already at 0.015~T/min (the magnetization jumps were detected only above 0.025 T/min \cite{2010_slobinsky}) stopping a field sweep exactly at the sound-velocity-peak position leads to a relaxation of the sound velocity (and the temperature) to the value characteristic of the valleys between the peaks (not shown), meaning that for much slower sweep rates the peaks should disappear (in agreement with Ref. \cite{2010_slobinsky}). This can also explain the broadening of the ultrasound features for the faster field sweeps when the relaxation time approaches the sweep time between the adjacent peaks (see Fig. \ref{fig2}).

Comparing different sweep rates unearthes a surprising feature: there appear to
exist two qualitatively different regimes. To illustrate
this statement we have plotted in the left panels of Fig. \ref{compare1} the
rising parts of the sound velocity peaks with different field-sweep rates (top)
and for different peaks (bottom) as a function of the applied field, while in
the right panels the descending parts of peaks and the temperature are plotted
as a function of time. We have plotted rising parts for the $c_T$
mode, and the descending one for the $c_L$ mode for different sweep rates,
for which the collapse is most prominent; the peak shapes of the two modes
behave a little different from each other. Scaling is evident in both
panels.

The explanation of the scalings is as follows. The external magnetic field 
drives the motion of a small number of thermally activated monopoles in
Dy$_2$Ti$_2$O$_7$ (which leads to a small response rate). For the system's 
state to keep up with the changing field, more monopoles need to be created 
so that spins can be flipped at a higher rate. However, energy barriers arise, 
from a competition between the Zeeman energy gain and the cost of creating 
and separating monopoles, and perhaps also from disorder.
%\cite{though}
Hence, the response has mostly intrinsic character in that it involves a 
distribution of magnetic energy barriers and not other degrees of freedom such 
as phonons. The energy released due to the growing number of monopoles moving 
along the field does not leave the sample quickly, leading to local heating 
(cf. Fig. \ref{fig2}) resulting in an avalanche \cite{2010_slobinsky}: More 
monopoles get thermally activated, more Zeeman energy is thus released etc.

On the other hand, the descending parts of the peaks in the sound velocity,
i.e., the release of the energy to the outside world, are different in
nature: The collapse of the curves as a function of time (regardless of field
strength) is consistent with the heat simply disappearing through a thermal
contact; indeed, the tail of the peak fits well to an exponential fall-off with
a timescale of roughly a minute. Crucially, the shape of the temperature traced
by the thermometer follows this peak decay quite closely, much more so than on
the rising side, where the temperature notably lags behind the sound
velocity's rise.

Two final remarks on the low-field regime are in order. First, our results
crisply complement neutron scattering \cite{2005_fennell} and the recent
magnetization \cite{2010_slobinsky} results on non-stationary field
sweeps, particularly by providing a detailed and time-resolved focus on the
nature of the thermal runaway and subsequent cool-down.

Second, very recent studies \cite{2010a_castelnovo,2011_giblin} of the monopole
dynamics in Dy$_2$Ti$_2$O$_7$ at low temperature suggest that there exist
several regimes of monopole relaxation in spin ice, in particular, with slow dynamics 
below $T_f$. From Fig. \ref{compare1}, we observe
that the relaxation of the sound velocity to the value characteristic of the
valleys between peaks is about a minute, in the same ballpark but somewhat shorter than a
monopole lifetime extracted in Ref.~\cite{2011_giblin}, $\sim 150$~s.
Notice, however, that relaxation times are observed to be field dependent
\cite{2011_klemke}.

Next, we turn to the strong-field regime, where there is a monopole liquid-gas
transition in spin ice. This takes place between the Kagome ice regime and the
saturated state, which have a low and high density of monopoles, respectively,
see Refs.~\cite{2008_castelnovo,2009_kadowaki,2006_tabata,2007_matsuhira} for
details. For our purposes it is enough to note that the magnetic field applied along the [111] direction acts as a staggered chemical potential for the monopoles
\cite{2008_castelnovo}. At low temperature, the monopole density exhibits the
discontinuous jump of a first-order transition, whereas there is a continuous
crossover at higher temperature. A critical endpoint separates the two \cite{2004_aoki}.

Measured values for the critical field, $H_c$, agree with the value $H_c \sim 1.25$~T
(see Fig. \ref{fig1}), observed via neutron scattering \cite{2007_fennell},
however, they disagree with $H_c \sim 0.93$ T, observed in magnetocaloric
and magnetization experiments \cite{2004_aoki,2003_sakakibara}. Anomalies are detected in specific-heat measurements at both $\sim 1$~T and $\sim 1.25$~T for $T < 0.3$~K \cite{2004_higashinaka}. A careful estimation of the demagnetization effect 
in our experiments leads to a corrected value of the $H_c =0.96\pm0.04$ T which is very close to one obtained in \cite{2004_aoki,2003_sakakibara}.  We also see an onset of the hysteresis, clearly visible at $T = 0.29$~K (Fig. 1)
but not at $T = 0.7$~K. In between, at $T = 0.4$~K, $\Delta v/v$ is very steep,
suggesting proximity to the critical endpoint.

In passing, we note that our results indicate that the exchange interaction in 
Dy$_2$Ti$_2$O$_7$ is not direct. This is suggested by the 
fact that the sound attenuation at $H_c$ is stronger in the transverse mode than in the 
longitudinal one, given that the leading contribution 
to the exchange-striction coupling is proportional to the scalar product of exchange 
gradient and the sound polarization \cite{1974_tachiki}.

In conclusion, we have studied nonequilibrium processes and the role of
magneto-acoustic interactions in spin-ice Dy$_2$Ti$_2$O$_7$ by means of
ultrasound measurements. Unusual anomalies have been detected in all acoustic modes studied. The correlation between magnetization plateaux and the peaks in
the sound velocity shows that the magneto-acoustic interactions can play
important role in our understanding of the spin-ice. We have found that our
ultrasound technique is particularly sensitive to these effects, where the
periodicity, shape, width, number, and distance between peaks can be crisply
extracted. The temperature of the sample follows the sound-velocity change at
the peak positions. Field-driven monopoles change the magnetic correlations in
the system, and phonons eventually carry the energy out of the system. We thus
observe a quasi-periodic change between thermodynamically unstable states
through non-equilibrium processes, like in the ``bottleneck'' effect. It will be interesting 
to contrast this in detail to the two-dimensional magnetic arrays \cite{2006_wang}, 
where the role of monopoles and avalanche processes in a non-thermal ensemble 
have been a recent focus of attention \cite{2009_moeller}.

\begin{acknowledgements}
A. A. Z. acknowledges the support from the Institute of Chemistry of V. N.~Karazin
Kharkov National University. R. M. is very grateful to the authors of
Refs.~\cite{2009_morris,2010_slobinsky} for collaboration on related work, and
to Claudio Castelnovo and Shivaji Sondhi for countless useful discussions. 
\end{acknowledgements}

%%%%%%%%%%%%%%%%%%%%%%%%%%%%%%%%%%%%%%%%%%%%%%%%%%%%%%%%%%%%%%%%%%%%%%%%%%%%%%

\end{document}